\documentclass[aps,pre,twocolumn,showpacs,superscriptaddress,groupedaddress]{revtex4}
\usepackage{graphicx}
\usepackage{color}
\usepackage{chemmacros}
\usepackage{filecontents}
\usepackage[titletoc,toc,title]{appendix}

\begin{document}

\title{Charge fluctuation effects on the shape of flexible polyampholytes with applications to Intrinsically disordered proteins}
\author{Himadri S. Samanta}\affiliation{Department of Chemistry, University of Texas at Austin, TX 78712}
\author{Debayan Chakraborty}\affiliation{Department of Chemistry, University of Texas at Austin, TX 78712}
\author{D. Thirumalai} 
\affiliation{Department of Chemistry, University of Texas at Austin, TX 78712}


\begin{abstract}
Random polyampholytes (PAs) contain positively and negatively charged monomers that are distributed randomly along the  polymer chain. The interaction between charges is assumed to be given by the Debye-Huckel potential.  We show that the size of the PA is determined by an interplay between electrostatic interactions, giving rise to the polyelectrolyte (PE) effect due to net charge per monomer ($\sigma$), and an effective attractive PA interaction due to charge fluctuations, $\delta \sigma$. The interplay between these terms gives rise to non-monotonic dependence of the radius of gyration, $R_g$ on the inverse Debye length, $\kappa$ when PA effects are important (${\frac{\delta \sigma}{\sigma}} > 1$). In the opposite limit, $R_g$ decreases monotonically with increasing $\kappa$.  Simulations of PA chains, using a charged bead-spring model, further corroborates our theoretical predictions. The simulations unambiguously show that conformational heterogeneity manifests itself among sequences that have identical PA parameters. A clear implication is that the phases of PA sequences, and by inference IDPs, cannot be determined using only the bare PA parameters ($\sigma$ and $\delta \sigma$). The theory is used to calculate the changes in $R_g$ on $N$, the number of residues for a set of Intrinsically Disordered Proteins (IDPs). For a certain class of IDPs, with $N$ between 24 to 441, the size grows as $R_g \sim N^{0.6}$, which agrees  with data from Small Angle X-ray Scattering (SAXS) experiments.

\end{abstract}
\date{\today}
\maketitle

\section{Introduction}
The shapes and dynamics of polyampholytes (PAs), which are polymers with monomers that carry both positive and negative charges, have been extensively studied \cite{Edwards80Ferroelectrics,Higgs91JCP,Srivastava96Mac,Barrat07ACP,Borukhov98EPJB,Dobrynin95JPF,Dobrynin04JPS,Lee00JCP}. Polyampholytes naturally occur in aqueous solution if the monomers contain acidic and basic groups. In this sense, all proteins are PAs in which charged residues are interspersed between hydrophobic and hydrophilic residues. Because of the simultaneous presence of positive and negative charges, the conformations of the PAs are determined by an interplay of electrostatic interactions, charge fluctuation effects (see below), as well as the stiffness of the backbone. In simple terms, we expect that repulsion between like-charges would stretch the chain whereas attraction would tend to make the polymer compact. Of course, in random PAs this balance is determined on an average performed over an ensemble of sequences (see below). If the number, $N$, of monomers is large then the PA is predicted to adopt compact conformations if the polymer is overall neutral (the number of positive and negative charges nearly cancel). On the other hand, if there is residual charge on the PA it is likely to be extended. It should be noted that there are differences in the behavior of the dependence of the radius of gyration, $R_g$, on $N$ depending on whether the the chain is globally neutral (plus and minus charges exactly cancel) or statistically neutral \cite{Yamakov00PRL} (residual charge when averaged over a large number of sequences scales as $\sqrt{N}$ with $N \gg 1$). Thanks to several insightful theoretical studies \cite{Higgs91JCP,Barrat07ACP,Srivastava96Mac,Gutin94PRE},  the complex phase behavior of PAs as a function of salt concentration and temperature have been elucidated. 

More recently, there has been renewed interest in PAs in the biophysics community because many eukaryotic proteins contain an unusually large fraction of charged residues \cite{Wright15NatMolCellBiol,vanderlee14ChemRev,Oldfield14ARBiochem}. As a consequence the favorable hydrophobic interactions cannot overcome the residual electrostatic interactions. For this reason, this class of proteins do not adopt globular structures unless it is in complex with another partner protein. Polypeptide sequences with this characteristic are referred to as intrinsically disordered proteins or IDPs because they do not have stable ordered structures under physiological conditions. It is also the case that there are protein sequences in which only  certain regions are disordered under nominal conditions. Because of the preponderance of such sequences and their roles in a variety of cellular functions and the potential role they play in diseases \cite{Dima04Bioinformatics,Oldfield14ARBiochem},  there is heightened interest in understanding their structural and dynamical properties \cite{Das15COSB,Zheng16JACS,Schuler16ARB,Levine17COSB}.  The IDPs, whose backbone is relatively flexible (persistence length in the range (0.6 - 1.0) nm), are low complexity sequences containing a large fraction of charged residues and smaller fraction of hydrophobic residues compared to their counterparts that adopt well-defined structures in isolation. As a consequence water is likely to be a good  or at best  a $\Theta$ solvent, which means that $R_g \approx N^{\nu}$ where $\nu$ is approximately 0.6 or 0.5. There are differences between IDPs and random PAs. (i) The sequences of IDPs are quenched, thus making it necessary to understand the conformations of a specific sequence. In other words, two sequences with identical charge composition could have drastically different structural characteristics. Of course, this could be the case for random PAs as well although this aspect has not been investigated as much. (ii) Unlike the case of PAs for which $N \gg 1$, which  allows one to develop analytical and scaling type arguments using well-developed methods in polymer physics,  typically studied IDPs have finite $N$, at best on the order of a few hundred residues. (iii) IDPs also contain uncharged amino acids, which are not usually considered when treating PAs using theory and simulations. Despite these differences, concepts from polyelectrolytes (PEs) and PAs have been used to envision the conformations of IDPs using the difference between positive and negative charge ($\sigma$) and net charge as appropriate variables \cite{Uversky00ProtSci}.     

The importance of sequence effects on the $R_g$ of PAs was first illustrated in a key note by Srivastava and Muthukumar \cite{Srivastava96Mac}. Using Monte Carlo simulations, with $N = 50$, they showed that there are substantial variations in $R_g$ in PAs (containing only charged monomers) for a globally neutral chain. This study showed that the location of charges (sequence specificity) plays a crucial role in determining the conformational properties.  More recently, Firman and Ghosh \cite{Firman18JCP} used Edwards model for charged polymers, encoding for the precise sequence in order to calculate $R_g$s for small $N$.  Their theory successfully accounted for simulations of synthetic IDPs \cite{Das13PNAS}, containing only a mixture of positive and negative charged residues. 

Here, we develop a theory to investigate the effects of charge fluctuations on the shapes of random PAs. In our model there is a probability, $p_+$, ($p_-$) that a monomer at location $s$ is  positively (negatively) charged. The probabilities, $p_+$ and $p_-$, should be calculated as follows.  The number of sequences, $M$, of a PA containing $N$ monomers is $M = 3^N$  because each monomer can either have a $+$ or a $-$ charge or is neutral. We assume that there are no correlations between charges along a given sequence, which implies that the magnitude of charge of monomer $s$ does not affect the value of $s^{\prime}$. Thus, $p_+$ and $p_-$ are independent of $s$.  Let $N_{+}(s)$ ($N_{-}(s)$) be the number of sequences with $+$  ($-$) charge at position $s$. Then, $p_+ = \frac{N_{+}(s)}{M}$ and  $p_- = \frac{N_{-}(s)}{M}$. Because $N_{+}(s) + N_{-}(s) + N_{0}(s) = M$ where  $N_{0}(s)$ is the number of sequences in which the $s^{th}$ monomer is neutral, the probability that the $s^{th}$ monomer in an ensemble of $M$ sequences is neutral is $1 - p_{+} - p_{-}$.


The  fluctuations in the ensemble of PA sequences arise because the normalized charge distribution is taken to be stochastic given by,
\begin{equation}
P[\sigma(s)] = p_+ \delta [\sigma(s) - 1] + p_{-} \delta [\sigma(s) + 1 ] + ( 1 - p_{+} - p_{-}) \delta [\sigma(s)].
\label{Dist}
\vspace{.2 in}
\end{equation}
The charges are measured in units of $e^-$. Because some monomers do not carry a charge (like in IDPs) $(p_{+} + p_{-}) \ne 1$. The mean $\langle \sigma(s) \rangle$ gives the net charge,  $p_{+} - p_{-}$, and the expression for the square of the charge fluctuations is, $\langle \delta \sigma^{2}(s) \rangle = p_{+} + p_{-} - (p_{+} - p_{-})^2$.  We refer to $\langle \sigma(s) \rangle$ and $\langle \delta \sigma (s)\rangle$, both of which are independent of $s$, as PA variables. We show that due to $\langle \delta \sigma^{2}(s) \rangle$ the $R_g$ is altered substantially, and could even induce a coil-globule transition even when the total charge on the PA is not globally neutral.  Because of the opposing behavior of polyelectrolyte  ($\sigma \ne 0$) and PA effects arising from charge fluctuations ($\langle \delta \sigma^{2}(s) \ne 0$), the dependence of $R_g$ on the Debye screening length could be non-monotonic. The phase diagram in the [$\langle \sigma(s) \rangle$,$\langle \delta \sigma(s) \rangle$] plane is rich.  We also apply the theory to calculate  $R_g$ of specific IDP sequences.  Remarkably, the theory reproduces quantitatively the $R_g$ values for the wild type Tau protein and various fragments obtained from the wild type Tau, which have been measured by Small Angle X-ray Scattering (SAXS) experiments \cite{Mylonas08Biochem}. In Tau, and other IDPs, charge fluctuations arise because of conformational heterogeneity, which we demonstrate explicitly elsewhere \cite{Upayan18JACS} for IDPs, and here for PAs using simulations. From now on we drop the angular brackets in both $\langle \sigma \rangle$ and $\langle \delta \sigma \rangle$.

\section{Theory}
We begin by considering the Edwards Hamiltonian for a polymer chain: 
\begin{equation}\label{hamiltonian}
\mathcal{H}=\frac{3k_B T}{2 a_0^2} \int\limits_0^N  \left(\frac{\partial \vec{r}}{\partial s}\right)^2 ds + k_B T{V}(\vec{r}(s)),
\end{equation}
where $\vec{r}(s)$ is the position of the monomer $s$, $a_0$ is the monomer size, $N$ is the number of monomers. The first term in the Eq.(\ref{hamiltonian}) accounts for chain connectivity, and the second term represents the sum of excluded volume interactions, electrostatic interactions, and effects of charge fluctuations (see below) due to the random values of charges in different positions in the ensemble of sequences.  The expression for ${V}(\vec{r}(s))$ is,
\begin{eqnarray}\label{Hapotential}
{V}(\vec{r}(s))&=&\frac{v_0}{(2\pi a_0^2)^{3/2}}\sum\limits_{s,s'=0}^{N}  
\text{exp}[{-\frac{(\vec{r}(s)-\vec{r}(s'))^2}{2a_0^2}}]\\ \nonumber &+&
 l_B \int_0^N \int_0^N ds~ds'~ \sigma(s) \sigma(s') \frac{e^{-\kappa \mid \vec{r}(s)-\vec{r}(s') \mid}}{\mid \vec{r}(s)-\vec{r}(s')\mid }\\ \nonumber
 &=&V_0 +V_1(\mid \vec{r}(s)-\vec{r}(s')\mid ).
\end{eqnarray} 


The first term in Eq.(\ref{Hapotential}) accounts for the non-specific two body excluded volume interactions. It differs insignificantly from the usual $\delta$ function potential used in the standard Edwards model. Of course, when $a_0$ is small compared to $R_g$, the precise form of this term is irrelevant, as long as it is short-ranged. In a good solvent ($v_0>0$), the polymer chain swells with $R_g \sim a_0 N^\nu$ $(\nu\approx 0.6)$, where as in a poor solvent ($v_0<0$), the size of the polymer is $R_g \sim a_0 N^\nu$ $(\nu\approx 1/3)$. Here, we consider a PA in a good solvent ($v_0>0$).

From Eq.(\ref{Hapotential}) one may obtain an effective interaction term between charges on the PA chain. By following the theory developed previously \cite{Ha97JPF}, we use the Hubbard-Stratonovich transformation to decouple the product of charges 
$\sigma(s)\sigma(s')$ in  Eq.\ref{Hapotential}. The partition function may be written as, 
\begin{eqnarray}\label{pf1}
Z=\mathcal{N}^{-1} &&\int d[\psi(\vec{r})] \text{exp}\left[-\frac{1}{2}\int d\vec{r}d\vec{r'} \psi(\vec{r})\right.\\ \nonumber &&
\left. V_1^{-1}(\mid \vec{r}(s)-\vec{r}(s')\mid )\psi(\vec{r'})\right]Z_\psi
\end{eqnarray}
where, $Z_\psi=\int d[\vec{r}]\text{exp}\left[ -V_0 - i\int ds\sigma(s) \psi(\vec{r}(s))\right] $, and 
$\mathcal{N}=\int d[\psi(\vec{r})] \text{exp}[-\frac{1}{2}\int d\vec{r}d\vec{r'} \psi(\vec{r}) V_1^{-1}(\mid \vec{r}(s)-\vec{r}(s')\mid )\psi(\vec{r'})]$.
If we assume that the charge distribution (Eq. \ref{Dist}) is annealed, it suffices to average $Z_\psi$ over the sequence of charges. 
With assumption that the charges $\sigma(s)$ at distant sites are not correlated, the partition function
averaged over sequence of charges to second order in $\psi$ becomes,~\cite{Ha97JPF}
\begin{eqnarray}\label{pf2}
<Z_\psi>_{seq}&=&\int\mathcal{D}[\vec{r}]\text{exp} \{-i\sigma\int\psi(\vec{r})c(\vec{r})d\vec{r} \\ \nonumber 
&-& \frac{1}{2} (\delta \sigma)^2 \int[\psi^2(\vec{r})-<\psi^2(\vec{r})>_\psi]c(\vec{r}) d\vec{r}
\end{eqnarray}
where the average value of the charge on the chain, $\sigma=<\sigma(s)>=p_+ -p_-$, the charge fluctuation, $(\delta \sigma)^2= <\sigma^2(s)-<\sigma(s)>^2>=p_+ + p_- -(p_+ -p_-)^2$ and the local monomer density, $c(\vec{r}) =\int ds \delta(\vec{r}(s)-\vec{r})$.
The term involving $(\delta \sigma)^2$, arising from the charge fluctuations, gives rise to the so called  PA effect, which is manifested as an effective attractive interaction of the screened Coulomb potential. Using Eq.(\ref{pf1}) and Eq.(\ref{pf2}), we perform the needed integration over $\psi(\vec{r})$ to obtain the following expression for the effective two body interaction term between charges on the PA,
\begin{eqnarray}\label{potential}
\mathcal{V}(\vec{r}(s))&=&\frac{v}{(2\pi a_0^2)^{3/2}}\sum\limits_{s,s'=0}^{N}  
\text{exp}[{-\frac{(\vec{r}(s)-\vec{r}(s'))^2}{2a_0^2}}]\\ \nonumber &+&
\sigma^2 l_B \int \int ds~ds'~  \frac{e^{-\kappa \mid \vec{r}(s)-\vec{r}(s') \mid}}{\mid \vec{r}(s)-\vec{r}(s')\mid }\\ \nonumber &-&
\frac{1}{2}(\delta \sigma)^4 l_B^2 \sum_{\{s,s'\}}~  \frac{e^{-2\kappa \mid \vec{r}(s)-\vec{r}(s') \mid}}{\mid \vec{r}(s)-\vec{r}(s')\mid^2 }.
\end{eqnarray} 
We neglect the three body interactions in the effective Hamiltonian in Eq.(\ref{potential}), which would be important if the PA were in a poor solvent.  In the work of Higgs and Joanny~\cite{Higgs91JCP}, the variational type calculation (see below) was done directly using Eq.~3. In this case, upon expansion to second order in ${V}(\vec{r}(s))$, the electrostatic potential (second term in Eq.~3)  generates a term $\propto \sigma(s) \sigma(s^{\prime}) \sigma(s^{\prime\prime})\sigma(s^{\prime\prime\prime})\sigma(s^{\prime\prime\prime\prime})$, which is random. When averaged over the ensemble of sequences, the coefficient of the third term is $\propto (p_+ + p_-)^{2}$ in \cite{Higgs91JCP}. In contrast,  we carry out averaging first as shown in Eq.~ (4), and hence obtain a different prefactor for the charge fluctuation ($\delta \sigma$) induced attraction term in Eq.~(6).


The screened Coulomb potential, the second term in Eq.(\ref{potential}), accounts for the interactions between charges separated by a distance $\mid \vec{r}(s)-\vec{r}(s')\mid$. The strength of the unscreened electrostatic interactions is characterized by the Bjerrum length $l_B=e^2/\epsilon k_B T$. The Debye screening length, $\kappa^{-1}$ determines the range of the electrostatic interactions.   
By changing the value of $\kappa$, and hence the range of charge interactions, the PA chain could undergo a coil-to-globule transition.   The value of $\kappa$ may be  the changed by decreasing or increasing the salt concentration.   The dimensionless parameter, $\sigma$, determines the net charge per residue on the polyelectrolyte chain. For a particular sequence, fraction $p=p_+ + p_-$ of the monomers are charged with the charge on each monomer being  $\pm e$. Therefore, the net charge per monomer is $\sigma=\mid p_+ - p_-\mid$. The third term in Eq.(\ref{potential}) is the attractive interaction term that is proportional to charge fluctuations ($\delta \sigma$). The PA affect arises due to the interaction between charge and dipoles formed between sequence of positive and negative charges. The charge-dipole interaction term decays as $\mid \vec{r}(s)-\vec{r}(s')\mid^{-2}$ and it is effectively screened (with a screening length $1/2\kappa$) due to the presence of other dipoles. In the absence of the third term the Hamiltonian would describe a polyelectrolyte, whose phases as function of temperature and $\kappa$ have been previous described using the methods used here \cite{Ha92PRA}.



In order to obtain $R_g$, we adopt the Edwards-Singh (ES) type variational calculation \cite{Edwards79JCSFT}, which has been extensively used in the polymer literature \cite{Muthukumar82JCP,Muthu87JCP,Higgs91JCP,Ha92PRA,Ha99JCP}.  More recently, the method was used to study sequence dependence of collapse of polypeptide chains \cite{Himadri17SM} and polyelectrolytes \cite{Firman18JCP} with application to a special class of synthetic IDPs.   In developing the theory, we assume that the interactions between charges exist only between specific monomers, described by the second and third term in Eq.(\ref{potential}). The sum is over the set of specific contacts between pairs $\{s_i, s_j\}$. We use the contact maps of IDP, generated in coarse-grained (CG) simulations of IDPs \cite{Upayan18JACS},  in order to assign the specific interactions.
The contact map from the simulation is computed by using a cutoff of 8 \AA.  The contacts are included between all side chain beads. In the two bead CG model,\cite{Upayan18JACS} the charges are positioned on the center of masses of the side chain beads, and therefore the contact map includes charge-charge contacts.  

The ES method is a variational type (referred to as the uniform expansion method) calculation that represents the exact Hamiltonian by a Gaussian chain with an effective monomer size, which is determined as follows.
Consider a virtual chain without excluded volume interactions, whose radius of gyration $\langle R_{g}^{2} \rangle=N a^{2}/6$ \cite{Edwards79JCSFT}, described by the Hamiltonian,
\begin{equation}
\mathcal{H}_v=\frac{3k_B T}{2 a^2} \int\limits_0^N  \left(\frac{\partial \vec{r}}{\partial s}\right)^2 ds.
\end{equation}
The monomer size in the trial Hamiltonian is $a$.
We split the deviation $\mathcal{W}$ between the virtual chain Hamiltonian and the real Hamiltonian as,
\begin{equation}
\mathcal{H}-\mathcal{H}_v=k_BT\mathcal{W}=k_BT(\mathcal{W}_1+\mathcal{W}_2),
\end{equation}
where
\begin{equation}
\mathcal{W}_1=\frac{3}{2 }\left(\frac{1}{a_0^2}-\frac{1}{a^2}\right) \int\limits_0^N  \left(\frac{\partial \vec{r}}{\partial s}\right)^2 ds, ~
\mathcal{W}_2=\mathcal{V}(\vec{r}(s)).
\end{equation}
The radius of gyration is $R_g^2=\frac{1}{N} \int\limits_0^N \langle\vec{r}^2(s)\rangle ds$, with the average being,
	$\langle\vec{r}^2(s)\rangle=\frac{\int r^2 e^{-\mathcal{H}_v/k_BT}e^{\mathcal{-W}} \delta\vec{r}}{\int e^{-\mathcal{H}_v/k_BT}e^{\mathcal{-W}} \delta\vec{r}}=\frac{\langle\vec{r}^2(s)e^{\mathcal{-W}}\rangle_v}{\langle e^{\mathcal{-W}}\rangle_v}$,
where, $\langle \cdots \rangle_v$ denotes the average over $\mathcal{H}_v$.

Assuming that the deviation $\mathcal{W}$ is small, we can calculate the average to first order in $\mathcal{W}$. The result is, 
$	\langle\vec{r}^2(s)\rangle \approx \frac{\langle\vec{r}^2(s)(1-\mathcal{W})\rangle_v}{\langle (1-\mathcal{W})\rangle_v} \approx \langle\vec{r}^2(s)(1-\mathcal{W})\rangle_v\langle (1+\mathcal{W})\rangle_v $
and the radius of gyration becomes,
\begin{eqnarray}\label{rg}
&&<R_g^2>=\frac{1}{N} \int\limits_0^N \langle\vec{r}^2(s)\rangle ds\\ \nonumber && = \frac{1}{N} \int\limits_0^N [\langle\vec{r}^2(s)\rangle_v + \langle\vec{r}^2(s)\rangle_v \langle\mathcal{W}\rangle_v -\langle\vec{r}^2(s)\mathcal{W}\rangle_v] ds.
\end{eqnarray}
If we choose the effective monomer size $a$ in $\mathcal{H}_v$, such that the first order correction (second and third terms in the right hand side of Eq.(\ref{rg})) vanishes, then the size of the chain is, $\langle R_{g}^{2} \rangle=N a^{2}/6$. This is an estimate of the exact $\langle R_g^2 \rangle$, and is only an approximation as we have neglected $\mathcal{W}^2$ and higher powers of $\mathcal{W}$. Thus, in the ES theory, we determine $a$ using Eq. (\ref{rg}),
\begin{equation}\label{first}
\vspace{-.1 in}
 \frac{1}{N} \int\limits_0^N [ \langle\vec{r}^2(s)\rangle_v \langle\mathcal{W}\rangle_v -\langle\vec{r}^2(s)\mathcal{W}\rangle_v] ds=0.
\end{equation}
 The equation above leads to a self-consistent equation for $a$, and is given by \cite{Edwards79JCSFT}:
\begin{equation}
\vspace{-.3 in}
 \frac{1}{a_0^2}-\frac{1}{a^2}=
\frac{  \frac{1}{N}\int\limits_0^N [\langle\vec{r}^2(s)\rangle_v \langle\mathcal{V}\rangle_v -\langle\vec{r}^2(s)\mathcal{V}\rangle_v]ds}{ \frac{a^2}{N}\int_{0}^N ds \ \langle\vec{r}^2(s)\rangle_v}.
\vspace{.2 in}
\end{equation}
By calculating the averages in the Fourier space ($\vec{r}_n=\frac{1}{N}\int\limits_1^N \cos\left({ \frac{\pi n s}{N}}\right) \vec{r}(s) ds$; $\vec{r}(s)=2\sum\limits_{n =1}^{N}\cos\left({\frac{\pi n s}{N}}\right)\vec{\tilde{{r}_n}}$; $R_g^2=2\sum\limits_n \langle|{\vec{\tilde{r_n}}}^2|\rangle$), we obtain
\begin{widetext}
\vspace{-.15 in}
\begin{eqnarray}\label{selfconsistent}
\frac{1}{a_0^2}-\frac{1}{a^2}&=&\frac{4N a_0^3}{9\pi \sum{\frac{1}{n^2}}}\sum\limits_{s,s'=0}^N  
    \frac{C^{ss'}_{1}}{(\frac{a^2 N}{3\pi^2}C^{ss'}_{2}+a_0^2)^{\frac{5}{2}}} \\ \nonumber &&
+  \frac{4N \sigma^2 l_B}{9\pi^3 \sum{\frac{1}{n^2}}}\sum\limits_{s,s'=0}^N C^{ss'}_{1}
\left( \frac{\pi^{1/2}(1-2\frac{2\kappa^2 a^2 N C^{ss'}_2}{3\pi^2})}{4(\frac{2a^2 N}{3\pi^2} C^{ss'}_{2})^{3/2}} +\frac{\pi \kappa^3}{2} e^{(\frac{2a^2 \kappa^2 N C^{ss'}_{2}}{3\pi^2})}\text{erfc}[ \kappa \sqrt{\frac{2a^2N}{3\pi^2} C^{ss'}_{2}}]\right)
\\ \nonumber 
&&- \frac{4N (\delta\sigma)^4 l_B^2}{9\pi^3 \sum{\frac{1}{n^2}}}\sum\limits_{\{s,s'\}} 
C^{ss'}_{1} \int_0^\infty dq~q^3 \left(\pi-\text{arctan}\left(\frac{2\kappa}{q}\right)\right)
\text{exp}\left(-q^2 \frac{2a^2N}{3\pi^2} 
C^{ss'}_{2}\right)
\end{eqnarray}
\end{widetext}
where, $C^{ss'}_{1}= \sum\limits_{n=1}^{N}\frac{1-\cos[n \pi(s-s')/N]}{n^4}$ and $C^{ss'}_2=\sum\limits_{n=1}^{N}\frac{1-\cos[n \pi(s-s')/N]}{n^2}$. In obtaining Eq. \ref{selfconsistent} we have used $v_0 = \frac{4 \pi a_0^3}{3}$ in Eq. \ref{Hapotential}

From Eq.(\ref{selfconsistent}), we can calculate the effective monomer size $a$, and hence the chain size $<R_g^2>=\frac{a^2 N}{6}$.
However, without having to solve Eq.(\ref{selfconsistent}) numerically, we can define the $\Theta$-like point, which signals the onset of a potential transition from a coil to globule state in the PA. At the $\theta$-point, the repulsive terms exactly balance the PA term. Since at the $\Theta$-point, the PA behaves as a Gaussian chain, with $a=a_0$, we substitute this value for $a$ in Eq.\ref{selfconsistent}  to determine the the condition for the $\Theta$-point.
Thus, from Eq.(\ref{selfconsistent}), the critical charge fluctuation value, at which the PA term equals the excluded volume and PE terms is,
\begin{widetext}
\begin{eqnarray}\label{deltasigma}
(\delta\sigma_{\theta}^2)^2&=&\left[\frac{4N a_0^3}{9\pi \sum{\frac{1}{n^2}}}\sum\limits_{s,s'=0}^N  
    \frac{C^{ss'}_{1}}{(\frac{a^2 N}{3\pi^2}C^{ss'}_{2}+a_0^2)^{\frac{5}{2}}} \right. \\ \nonumber && 
\left.+  \frac{4N \sigma^2 l_B}{9\pi^3 \sum{\frac{1}{n^2}}}\sum\limits_{s,s'=0}^N C^{ss'}_{1}
\left( \frac{\pi^{1/2}(1-2\frac{2\kappa^2 a^2 N C^{ss'}_2}{3\pi^2})}{4(\frac{2a^2 N}{3\pi^2} C^{ss'}_{2})^{3/2}} +\frac{\pi \kappa^3}{2} e^{(\frac{2a^2 \kappa^2 N C^{ss'}_{2}}{3\pi^2})}\text{erfc}[ \kappa \sqrt{\frac{2a^2N}{3\pi^2} C^{ss'}_{2}}]\right)\right]/
\\ \nonumber 
&&\left[ \frac{4N  l_B^2}{9\pi^3 \sum{\frac{1}{n^2}}}\sum\limits_{\{s,s'\}} 
C^{ss'}_{1} \int_0^\infty dq~q^3 \left(\pi-\text{Arctan}\left(\frac{2\kappa}{q}\right)\right)
\text{exp}\left(-q^2 \frac{2a^2N}{3\pi^2} 
C^{ss'}_{2}\right)\right]
\end{eqnarray}
\end{widetext}
The numerator in Eq.(\ref{deltasigma}) is a consequence of the repulsion containing excluded volume interactions and polyelectrolyte term. The denominator encodes the PA effect, determining the extent to which the size of the polymer changes due to charge fluctuations.
Using Eq.(\ref{deltasigma}), we obtain the dependence of $\delta \sigma_{\theta}$ on $N$. Scaling $n$ by $N$, it can be shown that $C_1^{ss'}\sim \frac{1}{N^2}$ and $C_2^{ss'}\sim 1$. From these result, we obtain, $\delta \sigma_\theta \sim \sqrt{N}$.  The implication is that for $N \gg 1$, charge fluctuations have to be extremely large to drive coil to globule transition unless the PA is globally neutral. Because even for statistically neutral PA, the PE term would not be irrelevant, we surmise that a genuine coil to globule transition transition may not be easily realizable in long PAs, which is in accord with the results in a previous study \cite{Yamakov00PRL}.  By implication our theory suggests that maximally compact IDPs would be difficult to obtain for generic IDP sequences if the fractions of + and - charged residues is on the order of (0.4 - 0.5). Of course, to establish the various conformations IDPs or PAs adopt as the nominal PA parameters and salt concentration ($\sigma$, $\delta \sigma$, and $\kappa$) are varied, will require performing detailed calculations as was previously done for polyelectrolytes \cite{Lee01Macromolecules}.

\section{Simulations }

\textbf{Model:} To provide further insights into some aspects of our theoretical predictions, and to highlight the nature of the heterogeneous ensembles that are sampled, we carried out simulations of PA chains, with $N=50$.  We consider sequences having the same net charge, $\sigma$, but different charge distributions to elucidate the role of sequence in determining the size of PAs.   The PA chain is modeled using a standard bead-spring model for charged polymers, with the total potential energy, $U_{tot}$, given by:

\begin{equation}
U_{tot} = U_{ch} + U_{exv} + U_{elec}.
\end{equation}
\noindent
Here, $U_{ch}$ describes the chain connectivity between the beads, and is modeled using the FENE potential: 

\begin{equation}
U_{ch} = \sum_{i}^{N_{bonds}}-0.5 k R_{0}^{2} \ln \left[  1 - \left( \frac{l_{i}-l_{0}}{R_{0}} \right) ^{2}         \right].
\end{equation}
\noindent
In Eq.~16, $k  = 20$\,kcal mol$^{-1}$ \AA$^{-2}$ denotes the spring constant; $l_{0} = 3.8$\,\AA \,\, is the equilibrium bond length between the connected PA beads; and $R_{0} = $\,2 \AA\,\, controls the maximum allowable deformation of the covalent bonds. 

The excluded volume interactions between pairs of beads are described by a truncated and shifted Lennard-Jones potential:

\begin{equation}
U_{exv} =  \sum _{i,j} ^{N_{pairs}} 4 \epsilon \left[\left( \frac{\sigma}{r_{ij}}\right)^{12}   -    \left( \frac{\sigma}{r_{ij}} \right) ^ {6}  + \frac{1}{4} \right]. 
\end{equation}

Based on previous work,~\cite{Kremer,PA_sim, PA_sim2} we set $\epsilon = k_{B}T$, and $\sigma = l_{0}$.  The pairwise interactions between the beads are ignored if the distance is greater than $2^{1/6}\sigma$.  This cutoff ensures that the excluded volume term is purely repulsive. 

The interactions between charged beads are taken into account via the screened Coulomb potential:
\begin{equation}
U_{elec} = \sum_{i,j}^{N_{charged}} \frac{q_{i}q_{j}}{2 \varepsilon r_{ij}}\exp^{-\kappa r_{ij}}
\end{equation}
In  Eq.~18, $\varepsilon$, and $\kappa$ are the inverse Debye length, and the dielectric constant, respectively.  We consider only unit charges, i.e., $q = \pm e$.

\textbf{Simulations:} The conformational space of each PA chain is explored using Langevin dynamics.  For each PA bead,  the stochastic equation of equation is given by: $m\bm{\ddot r}_{i} = -\gamma \bm{\dot r}_{i} + \bm{F}_{i} + \bm{g}_{i}$, where $m$ is the mass, $\bm{F}_{i}$ is the conservative force acting on each bead, and $\gamma_{i}$ is the drag coefficient. The Gaussian random force, $\bm{g}_{i}$,  satisfies $\langle \bm{g}_{i}(t) \bm{g}_{j}(t^{\prime})\rangle = 6k_{B}T \gamma \delta_{ij} \delta(t-t^{\prime})$.  The drag coefficient $\gamma$ is given by: $\gamma = m/\tau_{eff}$, where $\tau_{eff} = \sigma (m/\epsilon)^{1/2}$ is the effective time scale.  We used a variant of the velocity Verlet scheme~\cite{honeycutt_dt} to integrate the equations of motion, using a time step of $\Delta t = 0.01\,\tau_{eff}$.  Each simulation was carried out for  1.2 $\times$ 10$^{9}$ steps to ensure proper equilibration, and to obtain meaningful statistics.  

\textbf{Analysis:} Following  Eq.~6, we can estimate the charge fluctuations for each PA chain from simulations using an approximate expression: 
\begin{equation}
\langle \delta^{2}{U_{elec}} \rangle \approx \frac{(k_{B}T)^{2} \delta \sigma ^{4}_{c} l_{B}^{2}}{\langle R_{g} \rangle ^{2}},
\end{equation}

\noindent
where $\delta U_{elec} = U_{elec} - \langle U_{elec} \rangle$ is the  fluctuation in the electrostatic energy (Eq.~18) about its mean,  and $\delta \sigma_{c}$ denotes the charge fluctuation computed from the ensemble of sequence-specific conformations generated from simulations (see below). 

To characterize the structural heterogeneity of the PA ensembles, and to identify the most populated equilibrium conformations,  we carried out hierarchical clustering of the simulation trajectories using a pairwise distance metric, $D_{ij}$ defined as:

\begin{equation}
D_{ij} = \frac{1}{N_{p}} \sum _{a,b} \vert \left( r_{a,b} ^{i} - r_{a,b}^{j} \right) \vert,
\end{equation}

\noindent
where $r_{a,b}^{i}$ and $r_{a,b}^{j}$ are the pairwise distances between the PA beads $a$ and $b$, in snapshots $i$ and $j$, respectively. Distinct geometric clusters were identified using the Ward variance minimization algorithm,\cite{ward} as implemented within the \textit{scipy} module.  The hierarchical organization of conformations into distinct families were visualized in the form of dendrograms.

\section{Results:}
\textbf{Theoretical Predictions:} From Eq. \ref{potential} it is easy to show that that the size of the PA should be determined by ${\frac{\delta \sigma}{\sigma}}$, which can be written as $\sqrt{(1 - \sigma_A)/\sigma}$ where $\sigma_A = \frac{(p_+ - p_-)^2}{\sigma}$, which in the IDP literature is referred to as the charge asymmetry parameter. Fig. (1), displaying the dependence of the radius of gyration for a PA chain, with randomly distributed charges, on the screening length, 
 shows that $R_g$ changes non-monotonically as $\kappa$ increases when $\frac{\delta \sigma}{\sigma} = 10$.  In this charge-fluctuation dominated regime, the behavior  can be explained  by noting that at small values of  $\kappa$,  $R_g$ increases due to the PA term until $\kappa  l_b \approx 0.19$. In this range of $\kappa$, the effective attractions between monomers, the PA effect, decreases by adding ions to the solvent.  As a result the size of the chain increases. For the PA whose $R_g$ is shown in Fig. (1),  at $\kappa  l_B=0.19$ the PA and PE effects balance each other, and the chain becomes a random coil.  Upon further increase in $\kappa$,  the decrease in $R_g$ (but the chain is not a globule) is  due to the dominance of the PA term.  In the opposite limit when $\frac{\delta \sigma}{\sigma} = 0.1$, both the dimensions of the chain are dominated by the PE term, and $R_g$ decreases with increasing $\kappa$. We expect that at sufficiently large values of $\kappa l_B$ the consequences of PE and PA effects are negligible, and hence, $R_g$ would have the value expected for a Flory random coil ($\nu = 0.6$). Interestingly, these trends are qualitatively similar to experiments on two IDPs (N-terminal domain of HIV-1 integrase, and human prothymosin-$\alpha$)~\cite{Mueller-Spaeth10PNAS}.

\begin{figure}[h]
  \includegraphics[width=0.4\textwidth]{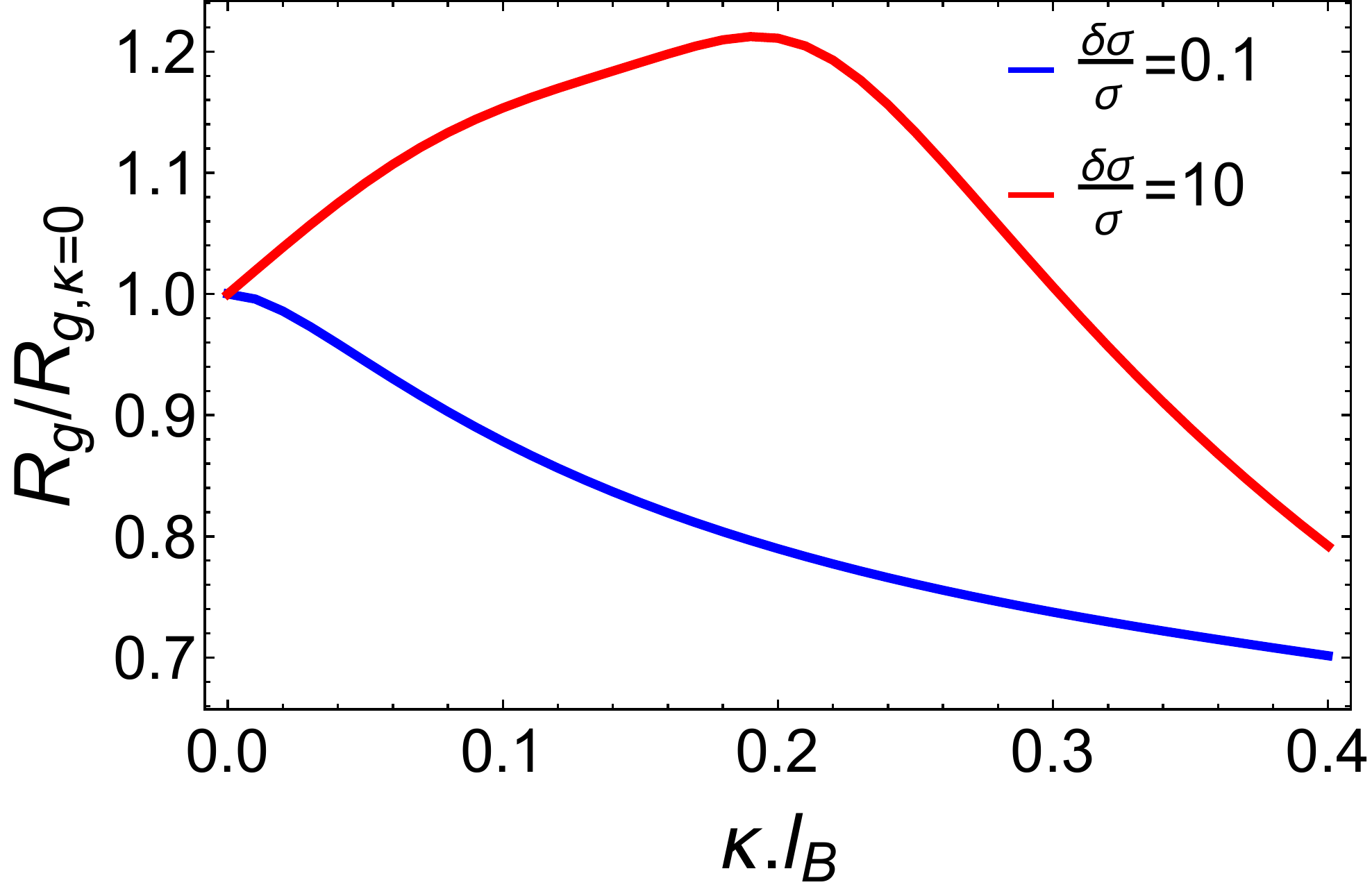}
\caption{Non-monotonic increase in the radius of gyration, $R_g$, with increasing value of the inverse screening length $\kappa$ for a PA chain (N=150).}
\end{figure}

\textbf{Predictions for a Tau-like IDP:} The generality of the theory allows us to predict the dependence size of an IDP. In Fig.(2) we plot  the $\kappa$ dependence of radius of gyration of a Tau protein fragment (K17Tau with $N$=145). 
To perform the calculations, we used the contact map generated in simulations based on the Self-Organized Polymer (SOP)-IDP model, which captures accurately the measured structure factors for a variety of IDPs \cite{Upayan18JACS}. Using input from simulations,~\cite{Upayan18JACS} which accounts for heterogeneity of the conformational ensembles of IDP, we find that $R_g$ of K17Tau changes non-monotonically with increasing value of $\kappa$ (Fig. (2)).   The size of K17Tau protein increases with $\kappa$ until it reaches a maximum at $\kappa l_B \approx 0.28$ (ion density for a monovalent salt is $\approx$ 30.5 mM), where the protein behaves like a polymer in a $\Theta$-solvent. The peak is broader with compared to the chain with random sequences. With further increase in $\kappa$, $R_g$ decreases just as for the random PA chain (Fig. 1). From these results, we conclude that charge fluctuations are substantial in K17Tau.   
\begin{figure}[h]
\vspace{.4 in}
 \includegraphics[width=0.4\textwidth]{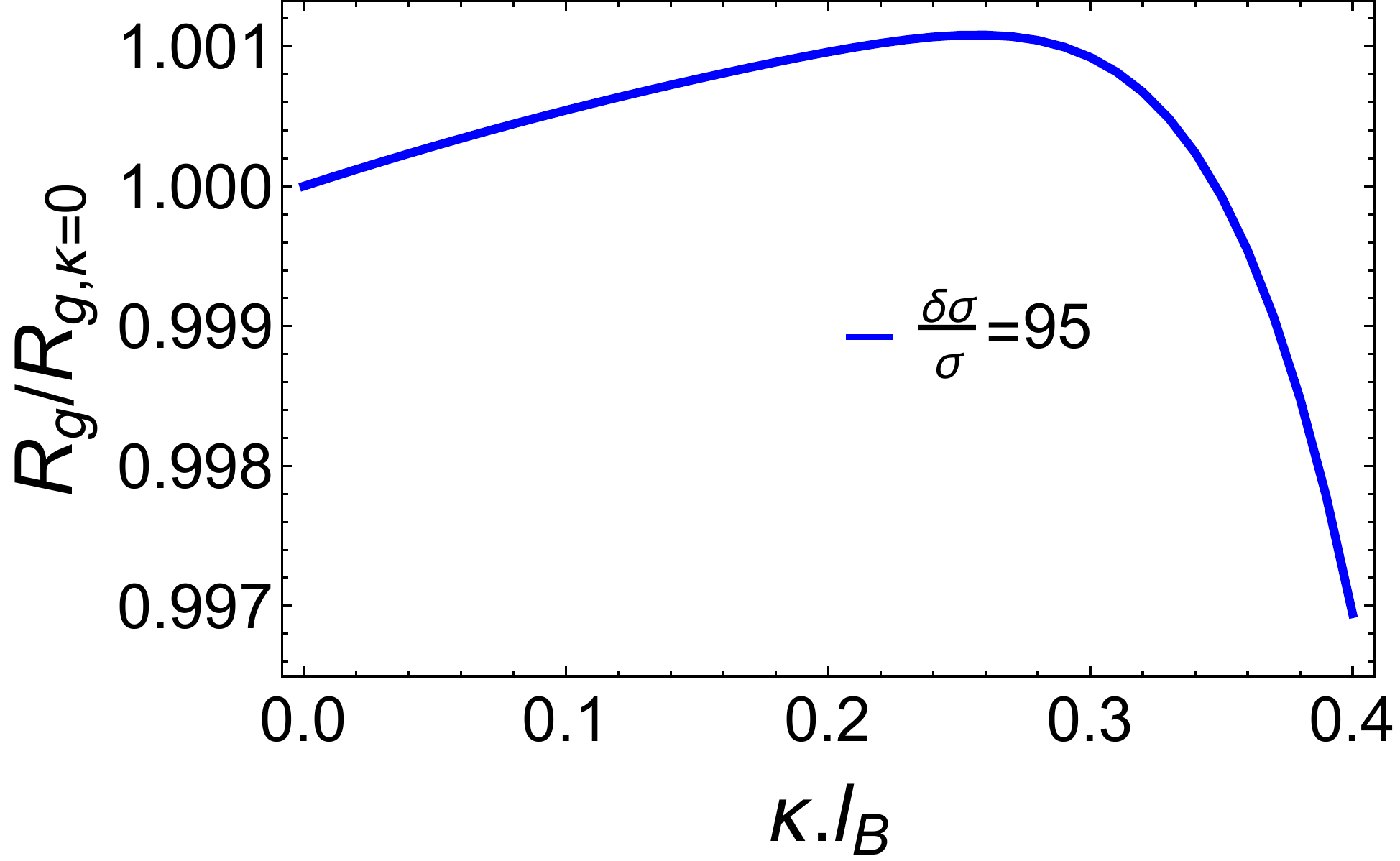}
\caption{The size of the chain ($R_g$) changes non-monotonically with increasing value of  $\kappa$ for K17Tau (N=145) protein. The parameter used to generate the plot are $\delta \sigma =0.95$ and $\sigma=0.01$.}
\end{figure}

\textbf{Phase diagram:} The 3D plots in fig.(3) and fig.(4) for two different values of $\kappa$ show the phase diagram  for different values of $\sigma$,   and $\delta \sigma$. The plot in fig.(3) shows that for small values of $\sigma$, the change in $R_g$ is significant at a particular value of $\delta \sigma$.  For a large value of net charge, say $\sigma=0.8$, the change in $R_g$ is small over a range of values of $\delta \sigma$ indicating that PE effects dominate. The value of $\delta\sigma_\theta$ increases with $\sigma$ for a PA chain. In fig.(4), for $\kappa=0.4$\,nm$^{-1}$ and for a high value of net charge $\sigma$, the change in $R_g$ is significant at a particular value of $\delta \sigma$ indicating that PA effects dominate. The phase diagrams show that by changing the salt concentrations, the sizes of random PAs can be altered dramatically.

\begin{figure}[h]
  \includegraphics[width=0.5\textwidth]{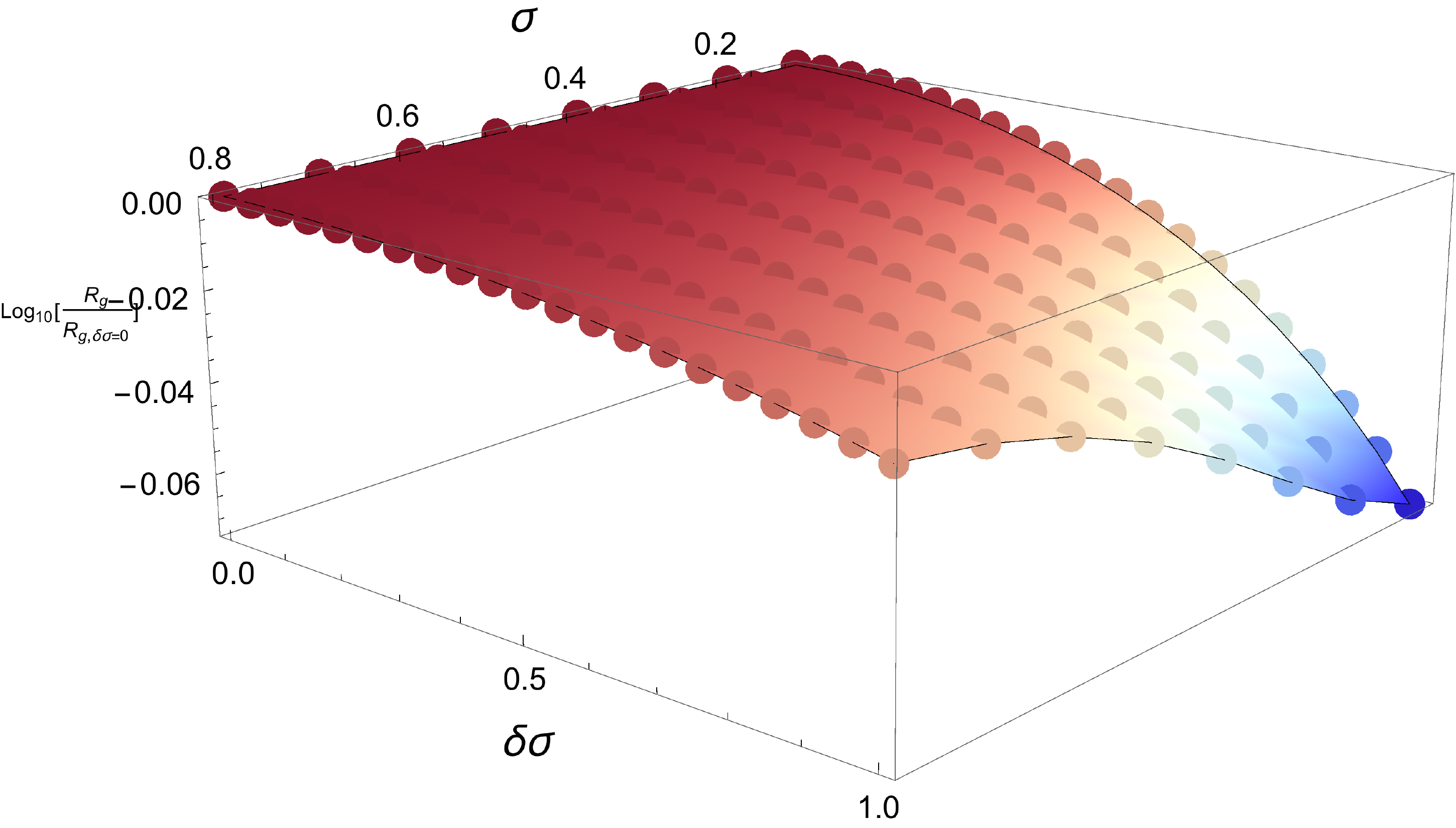}
\caption{Phase diagram for different values of net charge $\sigma$ and the charge fluctuation parameter $\delta \sigma$. The chain size decreases monotonically as  $\delta \sigma$ increases. The parameters for the plot: $N=150$ and $\kappa=0.2$\,nm$^{-1}$.}
\end{figure}
\begin{figure}[h]
  \includegraphics[width=0.5\textwidth]{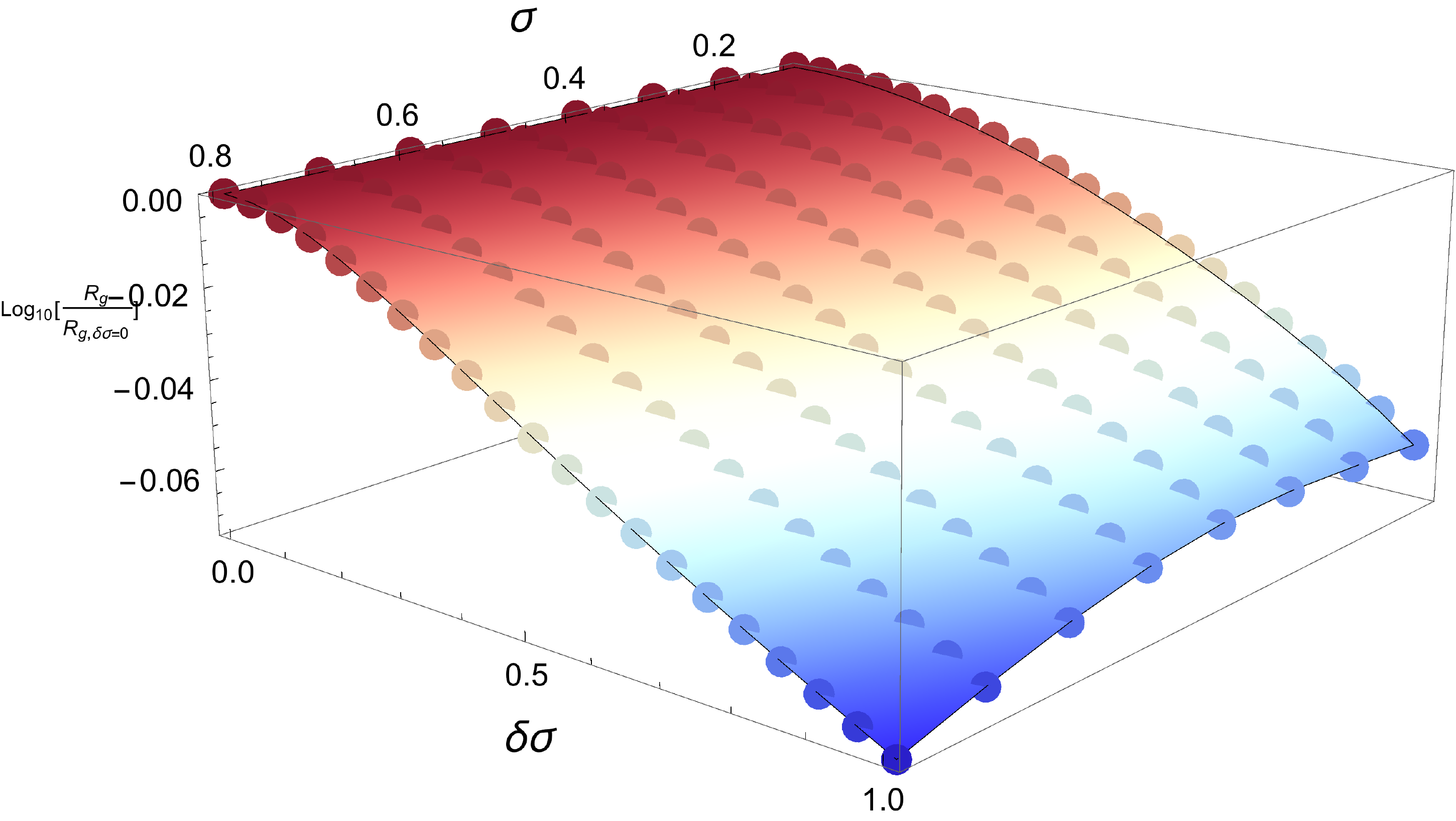}
\caption{Phase diagram for different values of net charge $\sigma$ and the charge fluctuation parameter $\delta \sigma$. The chain size decreases monotonically as  $\delta \sigma$ increases. The parameters for the plot: $N=150$ and $\kappa=0.4$\,nm$^{-1}$.}
\end{figure}

\textbf{Application to IDPs:} In order to calculate $R_g$ for several IDPs (see Fig. 5), with $N$ ranging from 24 (HISTATIN5) to 441 (hTau40) we used the average contact maps from simulations \cite{Upayan18JACS}, which restricts the summation in Eq.\ref{potential} to specific sites on the IDP. The dependence of $R_g$ on  the chain length for a set of IDPs (listed in the caption to Fig. 5) is shown in Fig. 5.  The theory shows that 
$R_g \sim N^{0.6}$, implying that these IDPs behave as self avoiding polymers, similar to the results in  the simulations for PAs \cite{Yamakov00PRL}.  The scaling in Fig. 5 has a weaker $N$ dependence than predicted by the renormalization group argument ($R_g \sim N$) for long PAs \cite{Kantor91EPL}.  The $N$ dependence in Fig. 5 has higher power than the result in  \cite{Higgs91JCP} ($R_g \sim N^{1/3}$).  It appears that for values of $\sigma$ observed in this set of IDPs the random coil behavior is the apt description. 


\begin{figure}[h]
  \includegraphics[width=0.4\textwidth]{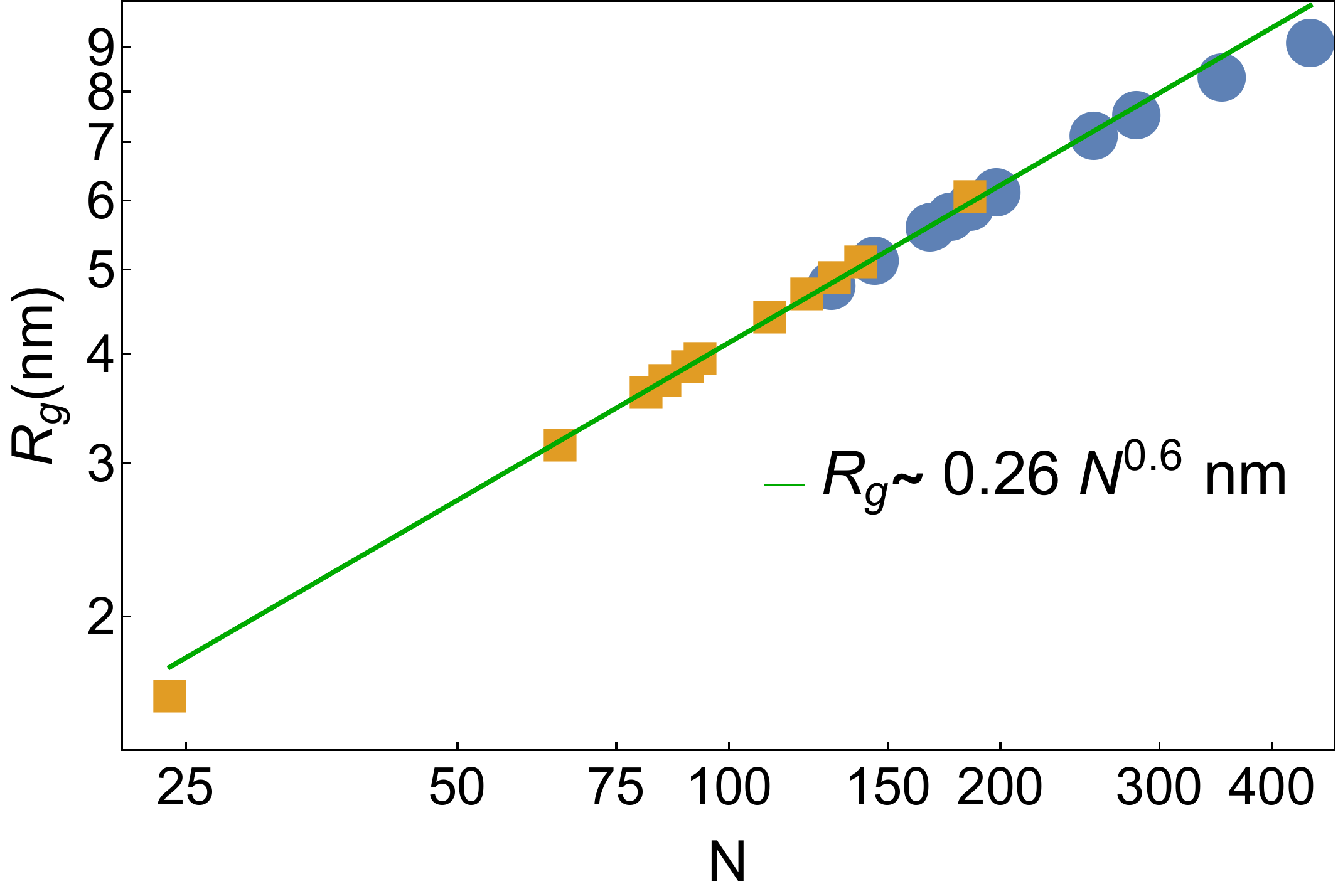}
  
\caption{The size $R_{g}$ increases with $N$ as $R_g \sim 0.26 N^{0.6}$ for specific contacts.  The symbols denote to the $R_g$ values calculated from the theory (using Eq.~6), and the green line is the power law fit.  The scaling is unchanged even when the restriction to specific interactions in Eq.~6 is removed.  The orange squares correspond to the $R_{g}$ values for the IDP sequences: ACTR (N=65), An16 (N=185), aSyn (N=140), ERMnTAD (N=122), HISTATIN5 (N=24), hNHE1 (N=131), NUP153 (N=81), p53 (N=93), ProtA (N=111), SH4UDsrc (N=85) and Sic1 (N=90).  The $R_g$ values for the hTau40 protein, and other constructs derived from it are denoted as blue circles. The various Tau protein constructs are: K18Tau ($N$=130), K17Tau ($N$=145), K27Tau ($N$=167), K10Tau ($N$=168), K32Tau ($N$=198), K44Tau ($N$=283), K23Tau ($N$=254), K25Tau ($N$=185), hTau23 ($N$=352), hTau40 ($N$=441) and K16Tau ($N$=176). The parameters used to generate the plot are $\sigma = 0.01$, $\delta \sigma =0.8$ and $\kappa l_B=0.324$. }

\end{figure}


\textbf{Sequence effects and conformational heterogeneity:} To illustrate the effect of charge fluctuations on the conformational ensemble of PAs we performed simulations of PA chains using a simple off-lattice model.  The PA variables, $p_{+}$, and $p_{-}$,  as well as the other simulation parameters are kept fixed. Hence, any variations in the size, or the underlying conformational heterogeneity of the PA sequences is entirely due to the different charge distributions.  In a recent study, Firman and Ghosh~\cite{Firman18JCP} identified combinations of $p_{+}$ and $p_{-}$, for which coil to globule transitions are expected to be extremely sensitive to the charge decoration along the PA chain.  Taking a cue from their work, we consider PA sequences with a net  charge of +6, with $p_{+} = 0.280$, and $p_{-} = 0.160$. 

Twenty different realizations of charge distributions were generated by randomly permuting the positions of the neutral, positive, and negative beads along the chain.  The ensemble averaged $R_{g}$ values fall in the range from 1.77 to 2.02\,nm.  The spread in $R_{g}$s is interesting considering that in all these sequences $N=50$, and the PA variables, $\sigma$ and $\delta \sigma$, which are often used to analyze data in the IDP community, are identical.  To explain these differences, in terms of fluctuations in the conformations, linked to $\delta \sigma$, we consider three representative examples (\textit{Seq1}, \textit{Seq2}, and \textit{Seq3}, in Fig.~6).  The peak of the $R_{g}$ distribution progressively shifts towards lower values in going from Seq1 to Seq3 (Fig.~6), clearly indicating that standard PA variables are not sufficient to fully describe the equilibrium properties. 

Insights into the relative populations of the coil-like and the globule-like  states for the three PA sequences can be obtained from the hierarchical arrangement of structural clusters (Fig.~7).  The structural ensemble of \textit{Seq1} is clearly dominated by extended conformations,  which accounts for 64.1\% of the equilibrium population.  Compact structures on the other hand, have a lower occupation probability (35.9\%).   For \textit{Seq2}, the relative populations of extended, and compact structures is approximately the same, being 49.5\%, and 50.5\%, respectively.  As is evident from the dendrogram, the equilibrium ensemble of \textit{Seq3} is primarily dominated by compact structures (net population of 73.4\%), which is in complete contrast to \textit{Seq1}.  The contrasting heterogeneity of the conformational ensembles for the three PA sequences, with identical $N$, $\sigma$, and $\delta \sigma$, readily explains the differences in $R_{g}$.

The distributions of $\delta U_{elec}$ (Fig.~6), together with the approximate values of $\delta \sigma_{c}$  computed using Eq.~19 (\mbox{Table 1}), provide a clear-cut evidence of the key role of the charge fluctuations in determining chain dimensions. For \textit{Seq1}, which has a high propensity to form extended structures, the $\delta U_{elec}$ distribution is narrow, and the charge fluctuation is minimal. In fact, the variance to mean ratio (VMR) of the electrostatic energy suggests that the charge distribution of \textit{Seq1} would correspond to the theoretical limit,$\frac{\delta \sigma}{\sigma} < 1$, where PE effects dominate. For \textit{Seq3}, the $\delta U_{elec}$ distribution is quite broad, and charge fluctuations effects are the most dominant, in perfect harmony with the clustering analysis, which revealed that \textit{Seq3} is mostly associated with compact structures.  Furthermore, the VMR suggests that in contrast to \textit{Seq1}, the appropriate theoretical limit would be $\frac {\delta \sigma}{\sigma} > 1$, the regime where PA effects dominate. \textit{Seq2}, for which the equilibrium populations of compact and extended structures are approximately equal, presents an interesting scenario. As expected, the estimated charge fluctuation, falls between the two extremities. The VMR $\approx$\,1 implies that the corresponding theoretical limit would be $\frac{\delta \sigma}{\sigma} \approx 1$.  Therefore, the random coil like behavior predicted from structural clustering, which manifests itself due to a balancing act of the PA and PE effects.

 We draw two important conclusions. (i) The scaling of $R_{g}$ with $N$, which for a certain class of IDPs, obey Flory scaling law can be used to accurately determine $R_{g}$ without the need for experiments or  simulations. The class of IDPs for which $R_{g}$ can be computed using \mbox{$R_{g} \sim N^{\nu}$} ($\nu \approx 0.6$) can be discerned using $\sigma$ and $\delta \sigma$. (ii) However, quantitative description of the equilibrium properties of IDPs as a function of salt concentration or denaturants requires a complete characterization of the conformational ensemble, as simulations explicitly demonstrate. The analyses of charge fluctuations show that $\delta \sigma$, which can be readily calculated for a specific sequence (in our simulations they are identical) is substantially modified when weighted by the energies associated with the conformational ensemble (denoted as $\delta \sigma_{c}$). Thus, only by understanding the details of the conformations can the properties of IDPs be correctly described. Alas, the average $R_{g}$ masks such subtle but important effects.


\begin{table}
\begin{tabular}{| l | l | l | l | l | l}
\hline
Seq & $\langle \delta ^{2} {U_{elec}} (kcal/mol) \rangle$ & $\langle R_{g} \rangle$ (nm) & $\delta \sigma_{c}$ & $ \vert \langle \delta^{2}{U_{elec}} \rangle/\langle U_{elec} \rangle \vert$\\
\hline
Seq1 & 0.23 & 2.03 & 1.53 & 0.83 \\
Seq2 & 1.04 & 1.87 & 2.14& 1.02\\
Seq3 & 4.07 & 1.78 & 2.93 & 9.49\\
\hline
\end{tabular}
\caption{ The values of charge fluctuation,$\delta \sigma_{c}$, for the different sequences.  Note that $\delta \sigma_{c}$ is computed from the ensemble of sequence-specific conformations using Eq.~19. Also shown are the variance to mean ratios (VMR) for the electrostatic energy.}
\end{table}






\begin{figure}[htbp!]
\includegraphics[width=0.44\textwidth]{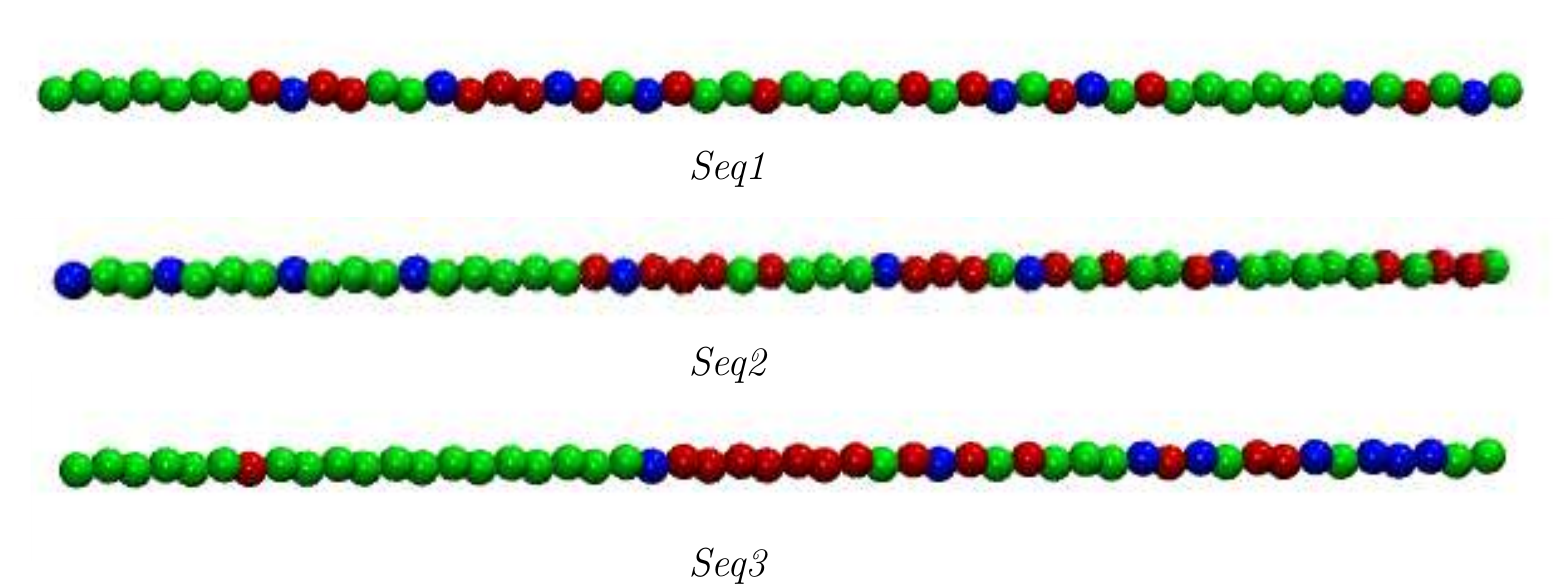}
\includegraphics[width=0.45\textwidth]{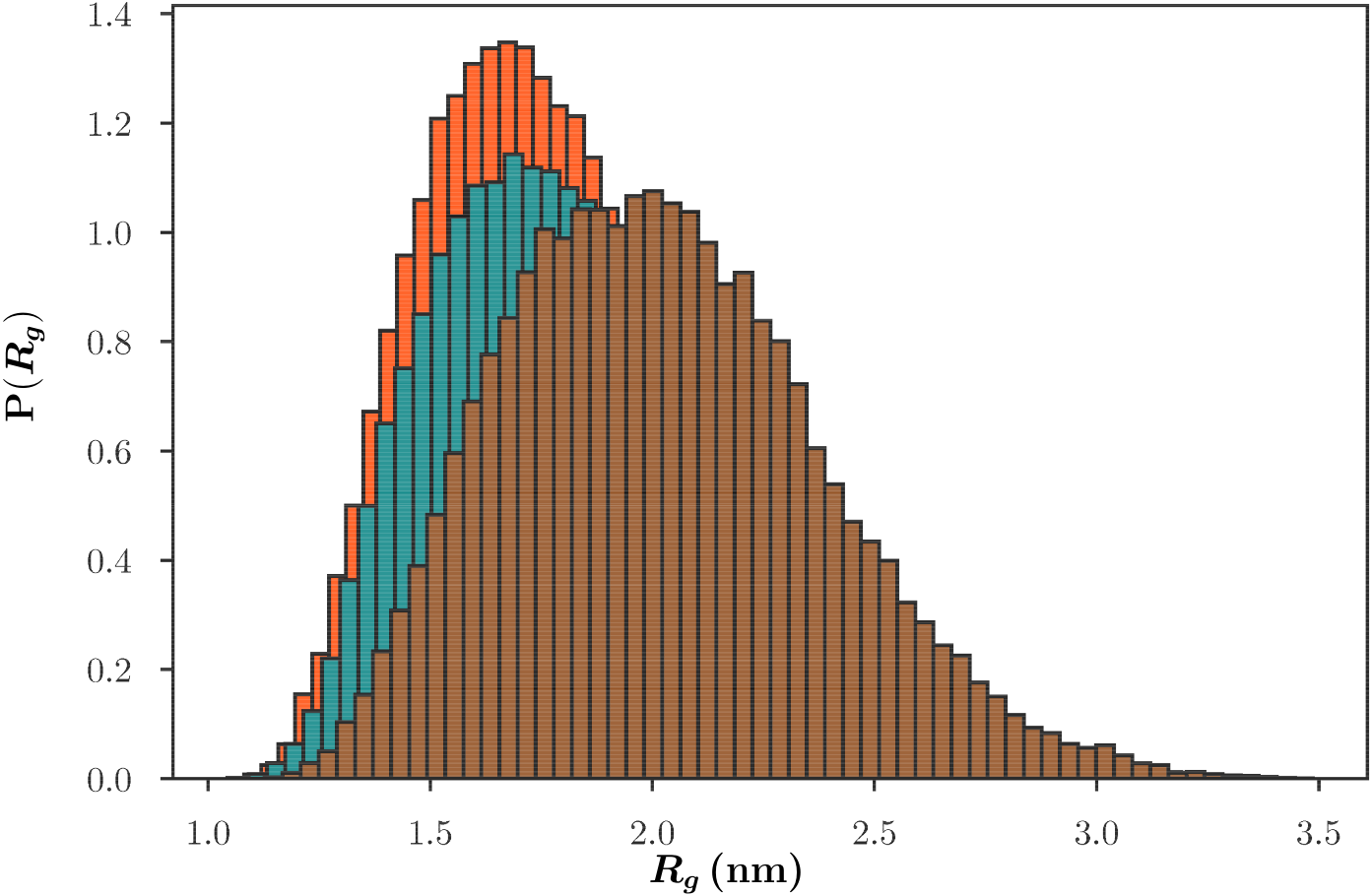}
\includegraphics[width=0.45\textwidth]{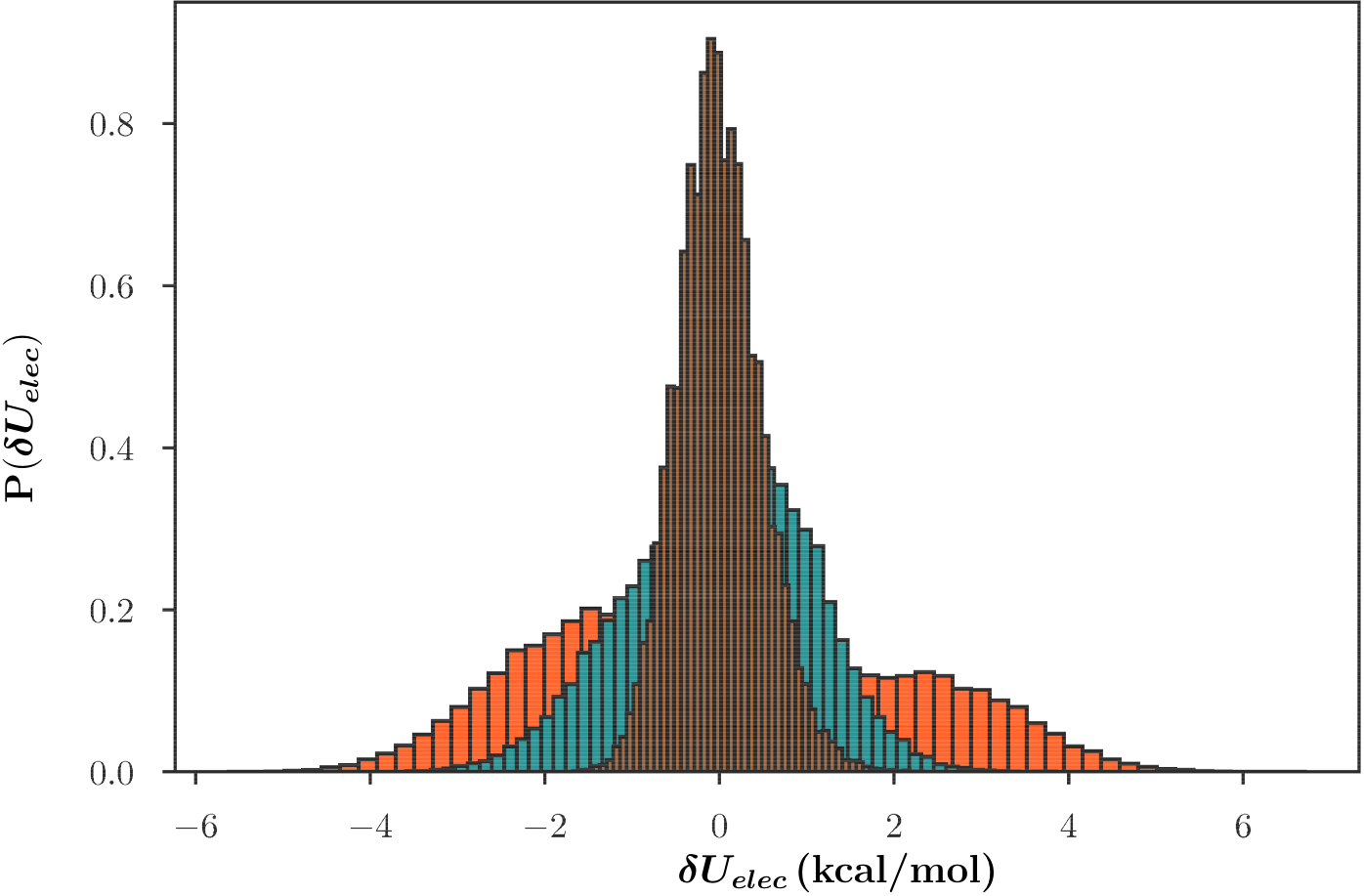}
\caption{Top: The cartoon representations of the PA sequences having different charge decorations along the chain with $p_{+}=0.280$, and $p_{-} = 0.160$. The beads are color coded according to charge: the neutral beads are colored green, positively charged beads are colored red, and negatively charged beads are colored blue. Middle: The distribution of $R_{g}$ for \textit{Seq1} (brown), \textit{Seq2} (cyan), and \textit{Seq3} (orange).  Bottom: The distributions of  $\delta U_{elec} = U_{elec} - \langle U_{elec} \rangle$ for the PA sequences, shown with the same color coding.}
\end{figure}

\begin{figure}[htbp!]
\includegraphics[width=0.44\textwidth]{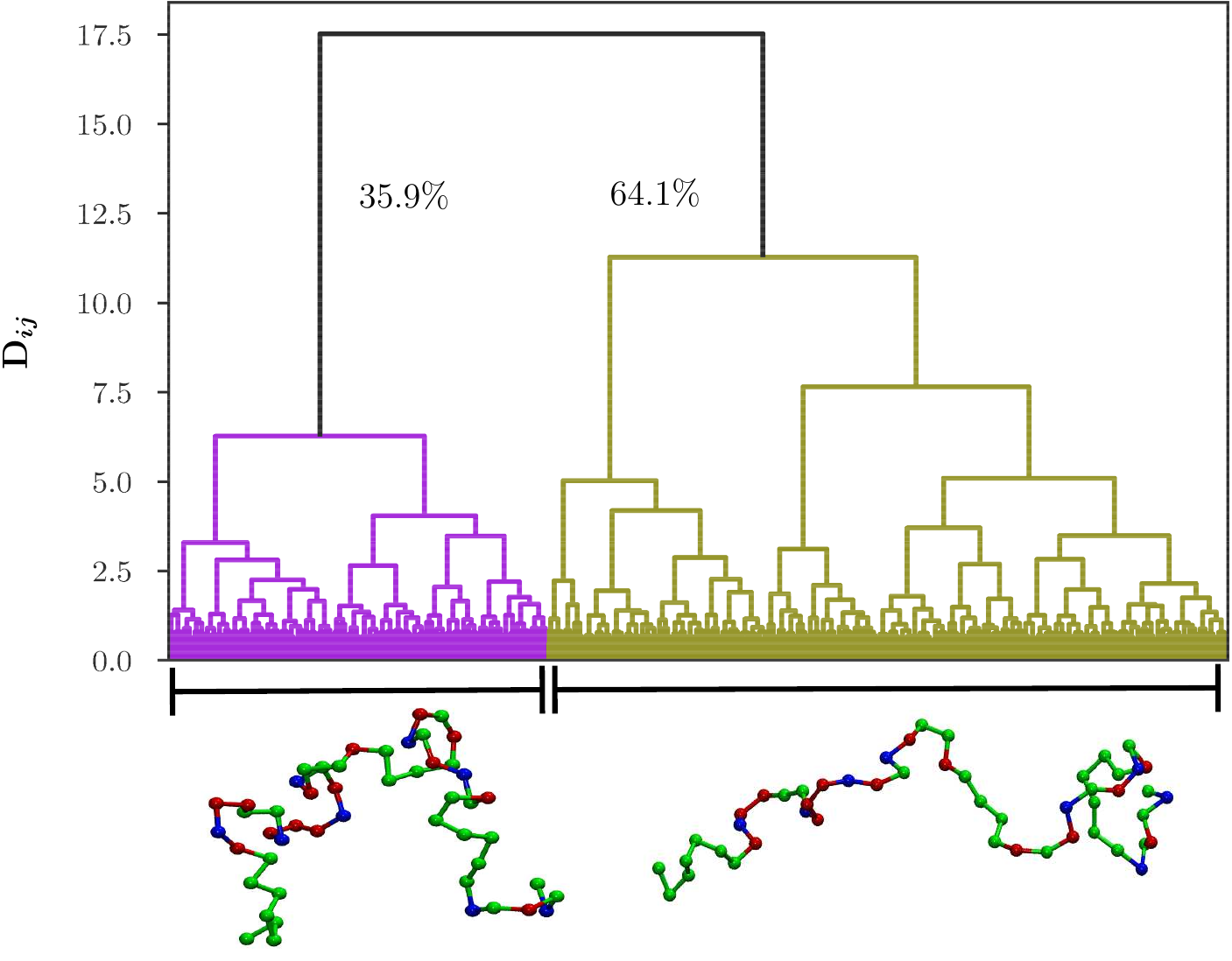}
\includegraphics[width=0.44\textwidth]{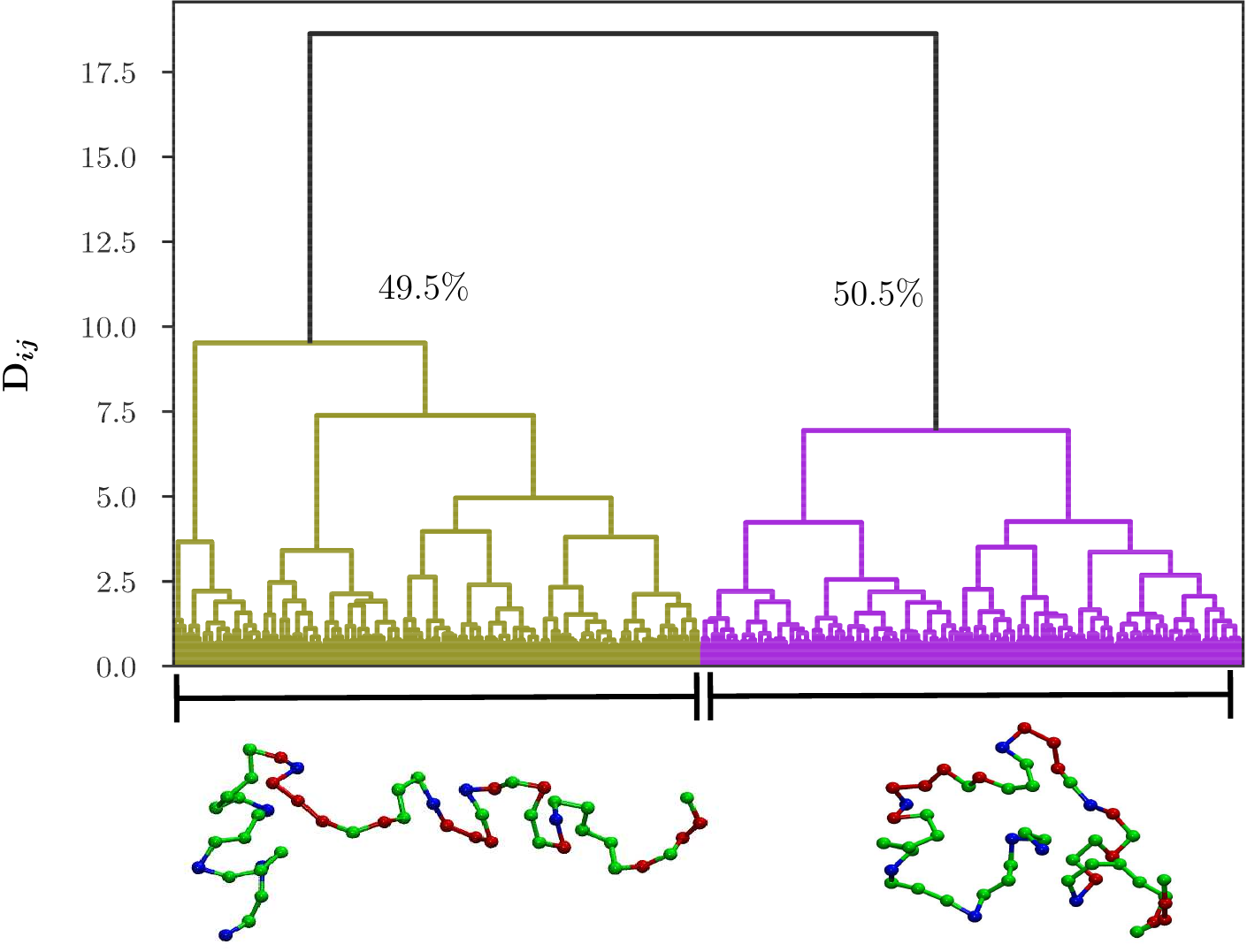}
\includegraphics[width=0.44\textwidth]{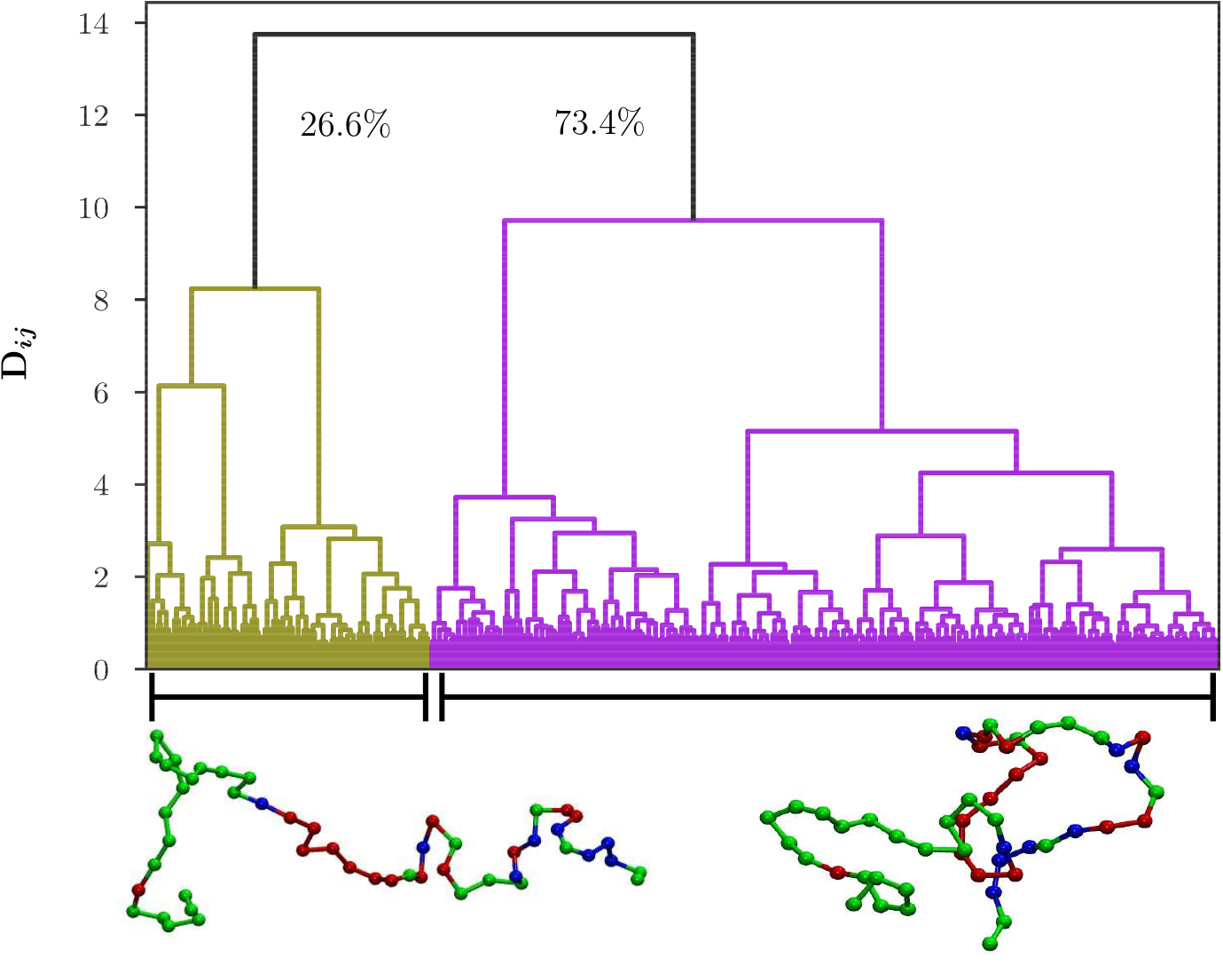}
\caption{The conformational heterogeneity of PA sequences depicted in the form of dendrograms (Top: \textit{Seq1}, Middle: \textit{Seq2}, and Bottom: \textit{Seq3}). The olive branches lead to extended configurations, and magenta branches lead to collapsed structures. Representative snapshots corresponding to the different clusters are also depicted.  The relative cluster populations are marked near the appropriate dendrogram branches.}
\end{figure}

\section{Conclusions:}
We developed a theory to quantitatively predict the effect of charge fluctuations ($\delta \sigma$)  on the size of flexible PAs that is in a good solvent (excluded volume interactions are positive) as a function of the inverse Debye length and  the net charge per monomer on the  chain $(\sigma$). Interestingly, when charge fluctuations are non-negligible ($\frac{\delta \sigma}{\sigma}$ is greater than unity), the radius of gyration increases non-monotonically as $\kappa$ increases. When $\frac{\delta \sigma}{\sigma}$ is less than unity, $R_g$ decreases with increasing $\kappa$. The generality of the theory allows us to predict $R_g$ for a number of IDPs. For a certain class of IDPs, we find the usual scaling of $R_g \sim N^{\nu}$ with $\nu = 0.6$, which coincides with the behavior expected for Flory random coils. Remarkably, our theory gives accurate estimates of the size of the Tau protein, and various fragments derived from it. This class of IDPs behaves as an ideal chain. The differences in the two scaling behavior between these IDPs can be rationalized in terms of the interplay between charge fluctuations and net charge per monomer.

What could be the origin of charge fluctuations in an IDP in which $\sigma$ (more precisely the precise sequence) is fixed? Even with  $\sigma$ fixed, the ensemble of conformations that a typical IDP or PA samples is heterogeneous. In sampling a large number of conformations, the spatial distances between charged residues could vary greatly. Therefore, the  effective charge of each conformation is different.  In some of the conformations, positively and negatively charged residues would be close together , whereas in others they would be spatially well-separated.  This gives rise to conformation-dependent effective attraction, which is quantified in our theory in terms of the average quantity $\delta \sigma$. Of course, the effective value of $\delta \sigma$ cannot be computed for a quenched charged sequence for a specific IDP without suitable simulations (see Figs 7 and 8, which illustrate this important point using three specific PA sequences). Therefore, it is difficult to construct phase diagrams of IDPs solely in terms of $\sigma$ or the differences between the number of positively and negatively charged residues. Construction of phase diagrams requires use of physical order parameters, which necessarily involves quantitatively characterizing the conformational ensembles of IDPs, an exercise requiring simulations using models that reproduce experimental measurements, such as, scattering profiles.   

{\bf Acknowledgments:} We are indebted to Upayan Baul for providing the simulation results and for useful discussions. We thank Prof. D. Svergun for providing us SAXS data on the Tau protein. This work was supported by the National Science Foundation (CHE 16-36424) and the Collie-Welch Foundation (F-0019).

\newpage

\begin{thebibliography}{10}

\bibitem{Edwards80Ferroelectrics}
S.~F. Edwards, P.~R. King, and P.~Pincus.
\newblock Phase changes in polyampholytes.
\newblock {\em Ferroelectrics}, 30(1):3--6, 1980.

\bibitem{Higgs91JCP}
P.~Higgs and J.-F. Joanny.
\newblock Theory of polyampholyte solutions.
\newblock {\em {J. Chem. Phys.}}, 94:1543, 1991.

\bibitem{Srivastava96Mac}
D.~Srivastava and M.~Muthukumar.
\newblock Sequence dependence of conformations of polyampholytes.
\newblock {\em Macromolecules}, 29:2324--2326, 1996.

\bibitem{Barrat07ACP}
J.‐L. Barrat and J.-F.~Joanny Joanny.
\newblock {\em Theory of Polyelectrolyte Solutions}, pages 1--66.
\newblock Wiley-Blackwell, 2007.

\bibitem{Borukhov98EPJB}
I.~Borukhov, D.~Andelman, and H.~Orland.
\newblock Random polyelectrolytes and polyampholytes in solution.
\newblock {\em Eur. Phys. J. B - Condensed Matter and Complex Systems},
  5(4):869--880, Nov 1998.

\bibitem{Dobrynin95JPF}
{A. V. Dobrynin} and {M. Rubinstein}.
\newblock Flory theory of a polyampholyte chain.
\newblock {\em J. Phys. II France}, 5(5):677--695, 1995.

\bibitem{Dobrynin04JPS}
A.~V. Dobrynin, R.~H. Colby, and M.~Rubinstein.
\newblock Polyampholytes.
\newblock {\em Journal of Polymer Science Part B: Polymer Physics},
  42(19):3513--3538, {2004}.

\bibitem{Lee00JCP}
N.~Lee and D.~Thirumalai.
\newblock Dynamics of collapse of flexible polyampholytes.
\newblock {\em {J. Chem. Phys.}}, 113(13):5126--5129, 2000.

\bibitem{Yamakov00PRL}
V.~Yamakov, A.~Milchev, H.~J. Limbach, B.~D$\ddot{\text{u}}$nweg, and
  R.~Everaers.
\newblock Conformations of random polyampholytes.
\newblock {\em {Phys. Rev. Lett.}}, 85:4305, 2000.

\bibitem{Gutin94PRE}
AM~Gutin and EI~Shakhnovich.
\newblock {Effect of a net charge on the conformation of polyampholytes}.
\newblock {\em {Phys. Rev. E}}, {50}({5}):{R3322--R3325}, {1994}.

\bibitem{Wright15NatMolCellBiol}
P.~E. Wright and H.~J. Dyson.
\newblock Intrinsically disordered proteins in cellular signalling and
  regulation.
\newblock {\em Nat. Rev. Mol. Cell Biology}, 16:18--29, 12 2015.

\bibitem{vanderlee14ChemRev}
M.~van~der Lee, R.and~Buljan, B.~Lang, R.~J. Weatheritt, G.~W. Daughdrill,
  A.~K. Dunker, M.~Fuxreiter, J.~Gough, J.~Gsponer, D.~T. Jones, P.~M. Kim,
  R.~W. Kriwacki, C.~J. Oldfield, R.~V. Pappu, P.~Tompa, V.~N. Uversky, P.~E.
  Wright, and M.~M. Babu.
\newblock {Classification of Intrinsically Disordered Regions and Proteins}.
\newblock {\em {Chem. Rev. }}, {114}({13}):{6589--6631}, {2014}.

\bibitem{Oldfield14ARBiochem}
C.~J. Oldfield and A.~K. Dunker.
\newblock {Intrinsically Disordered Proteins and Intrinsically Disordered
  Protein Regions}.
\newblock In {Kornberg, RD}, editor, {\em {Ann. Rev. Biochem.}}, volume~{83} of
  {\em {Annual Review of Biochemistry}}, pages {553--584}. {2014}.

\bibitem{Dima04Bioinformatics}
RI~Dima and D~Thirumalai.
\newblock {Proteins associated with diseases show enhanced sequence correlation
  between charged residues}.
\newblock {\em {Bioinformatics}}, {20}({15}):{2345--2354}, {2004}.

\bibitem{Das15COSB}
Rahul~K. Das, Kiersten~M. Ruff, and Rohit~V. Pappu.
\newblock {Relating sequence encoded information to form and function of
  intrinsically disordered proteins}.
\newblock {\em {Curr. Opin. Struct. Biol.}}, {32}:{102--112}, {2015}.

\bibitem{Zheng16JACS}
W.~Zheng, A.~Borgia, K.~Buholzer, A.~Grishaev, B.~Schuler, and R.~B. Best.
\newblock {Probing the Action of Chemical Denaturant on an Intrinsically
  Disordered Protein by Simulation and Experiment}.
\newblock {\em J. Am. Chem. Soc.}, {138}({36}):{11702--11713}, {2016}.

\bibitem{Schuler16ARB}
B.~Schuler, A.~Soranno, H.~Hofmann, and D.~Nettels.
\newblock {Single-Molecule FRET Spectroscopy and the Polymer Physics of
  Unfolded and Intrinsically Disordered Proteins}.
\newblock {\em {Ann. Rev. Biophys.}}, {45}:{207--231}, {2016}.

\bibitem{Levine17COSB}
Z.~A. Levine and J.-E. Shea.
\newblock {Simulations disordered proteins and systems with conformational
  heterogeneity}.
\newblock {\em {Curr. Opin. Struct. Biol.}}, {43}:{95--103}, {2017}.

\bibitem{Uversky00ProtSci}
V.~N. Uversky.
\newblock Natively unfolded proteins: A point where biology waits for physics.
\newblock {\em Protein Sci.}, 11:739--756, 2002.

\bibitem{Firman18JCP}
T.~Firman and K.~Ghosh.
\newblock Sequence charge decoration dictates coil-globule transition in
  intrinsically disordered proteins.
\newblock {\em {J. Chem. Phys.}}, 148:123305, 2018.

\bibitem{Das13PNAS}
R.~K. Das and R.~V. Pappu.
\newblock Conformations of intrinsically disordered proteins are influenced by
  linear sequence distributions of oppositely charged residues.
\newblock {\em {Proc. Natl. Acad. Sci.}}, 110:13392--13397, 2013.

\bibitem{Mylonas08Biochem}
E.~Mylonas, A.~Hascher, P.~Bernado, M.~Blackledge, E.~Mandelkow, and D.~I.
  Svergun.
\newblock {Domain conformation of tau protein studied by solution small-angle
  X-ray scattering}.
\newblock {\em {Biochemistry}}, {47}({39}):{10345--10353}, {2008}.

\bibitem{Upayan18JACS}
U.~Baul, D.~Chakraborty, M.~L. Mugnai, and D.~Thirumalai.
\newblock A sequence-specific two bead per residue model (sop-idp) for
  intrinsically disordered proteins.
\newblock {\em Unpublished}, 2018.

\bibitem{Ha97JPF}
B.-Y. Ha and D.~Thirumalai.
\newblock Persistence length of intrinsically stiff polyampholyte chains.
\newblock {\em J. Phys. II France}, 7:887--902, 1997.

\bibitem{Ha92PRA}
B.-Y. Ha and D.~Thirumalai.
\newblock Conformations of a polyelectrolyte chain.
\newblock {\em Phys. Rev. A}, 46:R3012--R3015, Sep 1992.

\bibitem{Edwards79JCSFT}
S.~F. Edwards and P.~Singh.
\newblock Size of a polymer molecule in solution. part 1.---excluded volume
  problem.
\newblock {\em J. Chem. Soc.{,} Faraday Trans. 2}, 75:1001--1019, 1979.

\bibitem{Muthukumar82JCP}
M.~Muthukumar and S.~F. Edwards.
\newblock Extrapolation formulas for polymer solution properties.
\newblock {\em The Journal of Chemical Physics}, 76(5):2720--2730, 1982.

\bibitem{Muthu87JCP}
M~Muthukumar.
\newblock {Adsorption of a polyelectrolyte chain to a charged surface }.
\newblock {\em {J. Chem. Phys.}}, {86}({12}):{7230--7235}, {1987}.

\bibitem{Ha99JCP}
B.-Y. Ha and D.~Thirumalai.
\newblock Persistence length of flexible polyelectrolyte chains.
\newblock {\em The Journal of Chemical Physics}, 110(15):7533--7541, 1999.

\bibitem{Himadri17SM}
H.~S. Samanta, P.~I. Zhuravlev, M.~Hinczewski, N.~Hori, S.~Chakrabarti, and
  D.~Thirumalai.
\newblock Protein collapse is encoded in the folded state architecture.
\newblock {\em Soft Matter}, 13:3622--3638, 2017.

\bibitem{Lee01Macromolecules}
N~Lee and D~Thirumalai.
\newblock {Dynamics of collapse of flexible polyelectrolytes in poor solvents}.
\newblock {\em {Macromolecules}}, {34}({10}):{3446--3457}, {2001}.

\bibitem{Kremer}
Gary~S. Grest and Kurt Kremer.
\newblock Molecular dynamics simulations of polymers in the presence of a heat
  bath.
\newblock {\em Phys. Rev. A}, 33:3628--3631, 1986.

\bibitem{PA_sim}
J.~Jeon and A.~V. Dobrynin.
\newblock Molecular dynamics simulations of polyampholyte−polyelectrolyte
  complexes in solutions.
\newblock {\em Macromolecules}, 38:5300--5312, 2005.

\bibitem{PA_sim2}
J.~Jeon and A.~V. Dobrynin.
\newblock Molecular dynamics simulations of polyelectrolyte−polyampholyte
  complexes. effect of solvent quality and salt concentration.
\newblock {\em J. Phys. Chem. B}, 110:24652--24665, 2006.

\bibitem{honeycutt_dt}
J.~D. Honeycutt and D.~Thirumalai.
\newblock The nature of folded states of globular proteins.
\newblock {\em Biopolymers}, 32:695--709, 1992.

\bibitem{ward}
J.~H. Ward.
\newblock Hierarchical grouping to optimize an objective function.
\newblock {\em Journal of the American Statistical Association}, 58:236--244,
  1963.

\bibitem{Mueller-Spaeth10PNAS}
S.~Mueller-Spaeth, A.~Soranno, V.~Hirschfeld, H.~Hofmann, S.~Rueegger,
  L.~Reymond, D.~Nettels, and B.~Schuler.
\newblock {Charge interactions can dominate the dimensions of intrinsically
  disordered proteins}.
\newblock {\em {Proc. Natl. Acad. Sci.}}, {107}({33}):{14609--14614}, {2010}.

\bibitem{Kantor91EPL}
Y.~Kantor and M.~Kardar.
\newblock Polymers with random self-interactions.
\newblock {\em Europhys. Lett.}, 14:421, 1991.

\end{thebibliography}


\end{document}